\documentclass[
 reprint,
superscriptaddress,
 amsmath,amssymb,
 aps
]{revtex4-2}
\usepackage{graphicx} 
\usepackage{dcolumn}
\usepackage{bm}
\usepackage{hyperref}
\usepackage[mathlines]{lineno}
\usepackage{xspace}

\makeatletter
\renewcommand*{\@fnsymbol}[1]{\ensuremath{\ifcase#1\or *\or \dagger\or \ddagger\or
   \mathsection\or \mathparagraph\or \|\or **\or \dagger\dagger
   \or \ddagger\ddagger \or \mathparagraph\mathparagraph \or \|\|  \else\@ctrerr\fi}}
\makeatother

\providecommand{\ssqthonethree}{\ensuremath{\sin^2\theta_{13}}\xspace}
\providecommand{\ssqthtwothree}{\ensuremath{\sin^2\theta_{23}}\xspace}

\providecommand{\ssqthtwoonethree}{\ensuremath{\sin^2 2\theta_{13}}\xspace}

\providecommand{\deltacp}{\ensuremath{\delta_{\scriptscriptstyle\mathrm{CP}}}\xspace}
\providecommand{\sindcp}{\ensuremath{\sin\deltacp}\xspace}

\providecommand{\Jcp}{\ensuremath{J_{\scriptscriptstyle\mathrm{CP}}}\xspace}

\providecommand{\dmsqtwothree}{\ensuremath{\Delta{}m^2_{32}}\xspace}

\providecommand{\nubar}{\ensuremath{\overline{\nu}}\xspace}
\providecommand{\nue}{\ensuremath{\nu_{e}}\xspace}
\providecommand{\numu}{\ensuremath{\nu_{\mu}}\xspace}

\providecommand{\nueb}{\ensuremath{\nubar_{e}}\xspace}
\providecommand{\numub}{\ensuremath{\nubar_{\mu}}\xspace}

\newlength{\parenbarKernelHeight}
\providecommand{\parenbar}[1]{
    \settoheight{\parenbarKernelHeight}{\ensuremath{#1}}%
    \addtolength{\parenbarKernelHeight}{0pt}
    \llap{\raisebox{\parenbarKernelHeight}{\scalebox{0.5}[0.5]{(}}}%
    \overline{#1}
    \rlap{\raisebox{\parenbarKernelHeight}{\scalebox{0.5}[0.5]{)}}}%
}
\providecommand{\nuany}{\ensuremath{\hspace{0.5mm}\parenbar{\nu}}\xspace}
\providecommand{\numuany}{\ensuremath{\nuany_\mu}\xspace}
\providecommand{\nueany}{\ensuremath{\nuany_{e}}\xspace}

\begin{document}

\title{First joint oscillation analysis of Super-Kamiokande atmospheric and T2K accelerator neutrino data 
}

\newcommand{\INSTBJ}{\affiliation{University of Tokyo, Institute for Cosmic Ray Research, Kamioka Observatory, Kamioka, Japan}} 
\newcommand{\INSTCG}{\affiliation{University of Tokyo, Institute for Cosmic Ray Research, Research Center for Cosmic Neutrinos, Kashiwa, Japan}} 
\newcommand{\INSTHA}{\affiliation{Kavli Institute for the Physics and Mathematics of the Universe (WPI), The University of Tokyo Institutes for Advanced Study, University of Tokyo, Kashiwa, Chiba, Japan}} 
\newcommand{\INSTHD}{\affiliation{University Autonoma Madrid, Department of Theoretical Physics, Madrid, Spain}} 
\newcommand{\INSTD}{\affiliation{University of British Columbia, Department of Physics and Astronomy, Vancouver, British Columbia, Canada}}
\newcommand{\INSTFE}{\affiliation{Boston University, Department of Physics, Boston, Massachusetts, U.S.A.}}
\newcommand{\INSTGA}{\affiliation{University of California, Irvine, Department of Physics and Astronomy, Irvine, California, U.S.A.}}
\newcommand{\AFFcsu}{\affiliation{California State University, Department of Physics, Dominguez Hills, Carson, California, U.S.A}} 
\newcommand{\AFFcnm}{\affiliation{Chonnam National University, Institute for Universe and Elementary Particles, Gwangju, Korea}} 
\newcommand{\INSTFH}{\affiliation{Duke University, Department of Physics, Durham, North Carolina, U.S.A.}}
\newcommand{\AFFgifu}{\affiliation{Gifu University, Department of Physics, Gifu, Japan}}
\newcommand{\AFFgist}{\affiliation{Gwangju Institute of Science and Technology, GIST College, Gwangju, Korea}} 
\newcommand{\AFFuh}{\affiliation{University of Hawaii, Department of Physics and Astronomy, Honolulu, Hawaii, U.S.A.}}
\newcommand{\INSTEI}{\affiliation{Imperial College London, Department of Physics, London, United Kingdom}}
\newcommand{\INSTCB}{\affiliation{High Energy Accelerator Research Organization (KEK), Tsukuba, Ibaraki, Japan}}
\newcommand{\INSTCC}{\affiliation{Kobe University, Kobe, Japan}}
\newcommand{\INSTCD}{\affiliation{Kyoto University, Department of Physics, Kyoto, Japan}}
\newcommand{\INSTFC}{\affiliation{University of Liverpool, Department of Physics, Liverpool, United Kingdom}}
\newcommand{\INSTCE}{\affiliation{Miyagi University of Education, Department of Physics, Sendai, Japan}}
\newcommand{\AFFnagoya}{\affiliation{Nagoya University, Institute for Space-Earth Environmental Research, Nagoya, Aichi, Japan}} 
\newcommand{\AFFkmi}{\affiliation{Nagoya University, Kobayashi-Maskawa Institute for the Origin of Particles and the Universe, Nagoya, Aichi, Japan}} 
\newcommand{\INSTDF}{\affiliation{National Centre for Nuclear Research, Warsaw, Poland}}
\newcommand{\INSTFJ}{\affiliation{State University of New York at Stony Brook, Department of Physics and Astronomy, Stony Brook, New York, U.S.A.}}
\newcommand{\INSTGJ}{\affiliation{Okayama University, Department of Physics, Okayama, Japan}}
\newcommand{\AFFosaka}{\affiliation{Osaka University, Department of Physics, Toyonaka, Osaka, Japan}}
\newcommand{\INSTGG}{\affiliation{Oxford University, Department of Physics, Oxford, United Kingdom}} 
\newcommand{\INSTE}{\affiliation{University of Regina, Department of Physics, Regina, Saskatchewan, Canada}}
\newcommand{\AFFseoul}{\affiliation{Seoul National University, Department of Physics, Seoul, Korea}} 
\newcommand{\INSTFB}{\affiliation{University of Sheffield, School of Mathematical and Physical Sciences, Sheffield, United Kingdom}}
\newcommand{\AFFshizuokasc}{\affiliation{Shizuoka University of Welfare, Department of Informatics in Social Welfare, Yaizu, Shizuoka, Japan}} 
\newcommand{\INSTEH}{\affiliation{STFC, Rutherford Appleton Laboratory, Harwell Oxford,  and  Daresbury Laboratory, Warrington, United Kingdom}}
\newcommand{\AFFskk}{\affiliation{Sungkyunkwan University, Department of Physics, Suwon, Korea}}
\newcommand{\INSTCH}{\affiliation{University of Tokyo, Department of Physics, Tokyo, Japan}}
\newcommand{\INSTHF}{\affiliation{Institute of Science Tokyo, Department of Physics, Tokyo, Japan}}

\newcommand{\INSTHG}{\affiliation{Tokyo University of Science, Faculty of Science and Technology, Department of Physics, Noda, Chiba, Japan}}
\newcommand{\INSTB}{\affiliation{TRIUMF, Vancouver, British Columbia, Canada}}
\newcommand{\AFFtokai}{\affiliation{Tokai University, Department of Physics, Hiratsuka, Kanagawa, Japan}}
\newcommand{\AFFtsinghua}{\affiliation{Tsinghua University, Department of Engineering Physics, Beijing,  China}} 
\newcommand{\INSTHE}{\affiliation{Yokohama National University, Department of Physics, Yokohama, Japan}}
\newcommand{\INSTBA}{\affiliation{Ecole Polytechnique, IN2P3-CNRS, Laboratoire Leprince-Ringuet, Palaiseau, France}}
\newcommand{\INSTGF}{\affiliation{INFN Sezione di Bari and Universit\`a e Politecnico di Bari, Dipartimento Interuniversitario di Fisica, Bari, Italy}}
\newcommand{\INSTBE}{\affiliation{INFN Sezione di Napoli and Universit\`a di Napoli, Dipartimento di Fisica, Napoli, Italy}}
\newcommand{\INSTBD}{\affiliation{INFN Sezione di Roma and Universit\`a di Roma ``La Sapienza'', Roma, Italy}}
\newcommand{\INSTBF}{\affiliation{INFN Sezione di Padova and Universit\`a di Padova, Dipartimento di Fisica, Padova, Italy}}
\newcommand{\INSTID}{\affiliation{Keio University, Department of Physics, Kanagawa, Japan}}
\newcommand{\INSTGH}{\affiliation{University of Winnipeg, Department of Physics, Winnipeg, Manitoba, Canada}}
\newcommand{\INSTIF}{\affiliation{King's College London, Department of Physics, Strand, London, United Kingdom}}
\newcommand{\INSTFD}{\affiliation{University of Warwick, Department of Physics, Coventry, United Kingdom}}
\newcommand{\AFFral}{\affiliation{Rutherford Appleton Laboratory, Harwell, Oxford, United Kingdom }}
\newcommand{\INSTDJ}{\affiliation{University of Warsaw, Faculty of Physics, Warsaw, Poland}}
\newcommand{\AFFbcit}{\affiliation{British Columbia Institute of Technology, Department of Physics, Burnaby, British Columbia, Canada }}
\newcommand{\INSTIJ}{\affiliation{Tohoku University, Faculty of Science, Department of Physics, Miyagi, Japan}}
\newcommand{\INSTHH}{\affiliation{Institute For Interdisciplinary Research in Science and Education (IFIRSE), ICISE, Quy Nhon, Vietnam}}
\newcommand{\INSTJD}{\affiliation{ILANCE, CNRS – University of Tokyo International Research Laboratory, Kashiwa, Chiba, Japan}}
\newcommand{\AFFibs}{\affiliation{Institute for Basic Science (IBS), Center for Underground Physics, Daejeon, Korea}}
\newcommand{\INSTHJ}{\affiliation{University of Glasgow, School of Physics and Astronomy, Glasgow, United Kingdom}} 
\newcommand{\AFFoecu}{\affiliation{Osaka Electro-Communication University, Media Communication Center, Neyagawa, Osaka, Japan}} 
\newcommand{\INSTJF}{\affiliation{University of Minnesota, School of Physics and Astronomy, Minneapolis, Minnesota, U.S.A.}} 
\newcommand{\INSTDI}{\affiliation{University of Silesia, Institute of Physics, Katowice, Poland}}

\newcommand{\INSTEE}{\affiliation{University of Bern, Albert Einstein Center for Fundamental Physics, Laboratory for High Energy Physics (LHEP), Bern, Switzerland}}
\newcommand{\INSTI}{\affiliation{IRFU, CEA, Universit\'e Paris-Saclay,  Gif-sur-Yvette, France}}
\newcommand{\INSTGB}{\affiliation{University of Colorado at Boulder, Department of Physics, Boulder, Colorado, U.S.A.}}
\newcommand{\INSTFG}{\affiliation{Colorado State University, Department of Physics, Fort Collins, Colorado, U.S.A.}}
\newcommand{\INSTJA}{\affiliation{E\"{o}tv\"{o}s Lor\'{a}nd University, Department of Atomic Physics, Budapest, Hungary}}
\newcommand{\INSTEF}{\affiliation{ETH Zurich, Institute for Particle Physics and Astrophysics, Zurich, Switzerland}}
\newcommand{\INSTIE}{\affiliation{CERN European Organization for Nuclear Research, Gen\'eve, Switzerland}}
\newcommand{\INSTEG}{\affiliation{University of Geneva, Section de Physique, DPNC, Geneva, Switzerland}}
\newcommand{\INSTDG}{\affiliation{H. Niewodniczanski Institute of Nuclear Physics PAN, Cracow, Poland}}
\newcommand{\INSTIB}{\affiliation{University of Houston, Department of Physics, Houston, Texas, U.S.A.}}
\newcommand{\INSTED}{\affiliation{Institut de Fisica d'Altes Energies (IFAE) - The Barcelona Institute of Science and Technology, Campus UAB, Bellaterra (Barcelona) Spain}}
\newcommand{\INSTJC}{\affiliation{Institut f\"ur Physik, Johannes Gutenberg-Universit\"at Mainz, Staudingerweg 7, Mainz, Germany}}
\newcommand{\INSTEC}{\affiliation{IFIC (CSIC \& University of Valencia), Valencia, Spain}}
\newcommand{\INSTEB}{\affiliation{Institute for Nuclear Research of the Russian Academy of Sciences, Moscow, Russia}}
\newcommand{\INSTHI}{\affiliation{International Centre of Physics, Institute of Physics (IOP), Vietnam Academy of Science and Technology (VAST), 10 Dao Tan, Ba Dinh, Hanoi, Vietnam}}
\newcommand{\INSTEJ}{\affiliation{Lancaster University, Physics Department, Lancaster, United Kingdom}}
\newcommand{\INSTII}{\affiliation{Lawrence Berkeley National Laboratory, Berkeley, California, U.S.A.}}
\newcommand{\INSTFI}{\affiliation{Louisiana State University, Department of Physics and Astronomy, Baton Rouge, Louisiana, U.S.A.}}
\newcommand{\INSTIH}{\affiliation{Joint Institute for Nuclear Research, Dubna, Moscow Region, Russia}}
\newcommand{\INSTHB}{\affiliation{Michigan State University, Department of Physics and Astronomy,  East Lansing, Michigan, U.S.A.}}
\newcommand{\INSTCF}{\affiliation{Osaka Metropolitan University, Department of Physics, Osaka, Japan}}
\newcommand{\INSTIC}{\affiliation{University of Pennsylvania, Department of Physics and Astronomy,  Philadelphia, Pennsylvania U.S.A.}}
\newcommand{\INSTGC}{\affiliation{University of Pittsburgh, Department of Physics and Astronomy, Pittsburgh, Pennsylvania, U.S.A.}}
\newcommand{\INSTFA}{\affiliation{Queen Mary University of London, School of Physics and Astronomy, London, United Kingdom}} 
\newcommand{\INSTGD}{\affiliation{University of Rochester, Department of Physics and Astronomy, Rochester, New York, U.S.A.}}
\newcommand{\INSTHC}{\affiliation{Royal Holloway University of London, Department of Physics, Egham, Surrey, United Kingdom}}
\newcommand{\INSTBC}{\affiliation{RWTH Aachen University, III. Physikalisches Institut, Aachen, Germany}}
\newcommand{\INSTJB}{\affiliation{Universidad de Sevilla, Departamento de F\'isica At\'omica, Molecular y Nuclear, Sevilla, Spain}}
\newcommand{\INSTBB}{\affiliation{Sorbonne Universit\'e, CNRS/IN2P3, Laboratoire de Physique Nucl\'eaire et de Hautes Energies (LPNHE), Paris, France}}
\newcommand{\INSTGI}{\affiliation{Tokyo Metropolitan University, Department of Physics, Tokyo, Japan}}
\newcommand{\INSTF}{\affiliation{University of Toronto, Department of Physics, Toronto, Ontario, Canada}}
\newcommand{\INSTDH}{\affiliation{Warsaw University of Technology, Institute of Radioelectronics and Multimedia Technology, Warsaw, Poland}}
\newcommand{\INSTEA}{\affiliation{Wroclaw University, Faculty of Physics and Astronomy, Wroclaw, Poland}}
\newcommand{\INSTH}{\affiliation{York University, Department of Physics and Astronomy, Toronto, Ontario, Canada}}
\newcommand{\INSTJE}{\affiliation{South Dakota School of Mines and Technology, Rapid City, South Dakota, U.S.A.}}
\newcommand{\INSTIG}{\affiliation{VNU University of Science, Vietnam National University, Hanoi, Vietnam}}

\newcommand{\INSTJG}{\affiliation{Ghent University, Department of Physics and Astronomy, Gent, Belgium}}

\INSTBJ
\INSTCG
\INSTHD
\INSTFE
\AFFbcit
\INSTGA
\AFFcsu
\AFFcnm
\INSTFH
\INSTBA
\AFFgifu
\AFFgist
\INSTHJ
\AFFuh
\AFFibs
\INSTHH
\INSTEI
\INSTGF
\INSTBE
\INSTBF
\INSTBD
\INSTJD
\INSTID
\INSTCB
\INSTIF
\INSTCC
\INSTCD
\INSTFC
\INSTJF
\INSTCE
\AFFnagoya
\AFFkmi
\INSTDF
\INSTFJ
\INSTGJ
\AFFoecu
\INSTGG
\AFFral
\AFFseoul
\INSTFB
\AFFshizuokasc
\INSTDI
\INSTEH
\AFFskk
\INSTIJ
\AFFtokai
\INSTCH
\INSTHA
\INSTHF
\INSTHG
\INSTB
\AFFtsinghua
\INSTDJ
\INSTFD
\INSTGH
\INSTHE

\author{K.~Abe}
\INSTBJ
\INSTHA
\author{S.~Abe}
\INSTBJ
\author{C.~Bronner}
\INSTBJ
\author{Y.~Hayato}
\INSTBJ
\INSTHA
\author{K.~Hiraide}
\INSTBJ
\INSTHA
\author{K.~Hosokawa}
\INSTBJ
\author{K.~Ieki}
\author{M.~Ikeda}
\INSTBJ
\INSTHA
\author{J.~Kameda}
\INSTBJ
\INSTHA
\author{Y.~Kanemura}
\author{R.~Kaneshima}
\author{Y.~Kashiwagi}
\INSTBJ
% ...


\collaboration{The Super-Kamiokande Collaboration}
\noaffiliation

\INSTHD
\INSTEE
\INSTFE
\INSTD
\INSTGA
\INSTI
\INSTGB
\INSTFG
\INSTFH
\INSTJA
\INSTEF
\INSTIG
\INSTIE
\INSTEG
\INSTHJ
\INSTJG
\INSTDG
\INSTCB
\INSTIB
\INSTED
\INSTJC
\INSTEC
\INSTHH
\INSTEI
\INSTGF
\INSTBE
\INSTBF
\INSTBD
\INSTEB
\INSTHI
\INSTJD
\INSTHA
\INSTID
\INSTIF
\INSTCC
\INSTCD
\INSTEJ
\INSTII
\INSTBA
\INSTFC
\INSTFI
\INSTIH
\INSTHB
\INSTCE
\INSTDF
\INSTFJ
\INSTEH
\INSTGJ
\INSTCF
\INSTGG
\INSTIC
\INSTGC
\INSTFA
\INSTE
\INSTGD
\INSTHC
\INSTBC
\INSTJF
\INSTJB
\INSTFB
\INSTDI
\INSTBB
\INSTJE
\INSTCH
\INSTBJ
\INSTCG
\INSTHF
\INSTGI
\INSTHG
\INSTF
\INSTB
\INSTDJ
\INSTDH
\INSTIJ
\INSTFD
\INSTGH
\INSTEA
\INSTHE
\INSTH

\author{K.\,Abe}\INSTBJ
\author{S.\,Abe}\INSTBJ
\author{N.\,Akhlaq}\INSTFA
\author{R.\,Akutsu}\INSTCB
\author{H.\,Alarakia-Charles}\INSTEJ
\author{A.\,Ali}\INSTGH\INSTB
\author{Y.I.\,Alj Hakim}\INSTEI
\author{S.\,Alonso Monsalve}\INSTEF
\author{C.\,Andreopoulos}\INSTFC
\author{L.\,Anthony}\INSTEI
% ...

\collaboration{The T2K Collaboration}\noaffiliation

\date{\today}

\begin{abstract}
The Super-Kamiokande and T2K collaborations present a joint measurement of neutrino oscillation parameters from their atmospheric and beam neutrino data. It uses a common interaction model for events overlapping in neutrino energy and correlated detector systematic uncertainties between the two datasets, which are found to be compatible. Using 3244.4 days of atmospheric data and a beam exposure of $19.7(16.3) \times 10^{20}$ protons on target in (anti)neutrino mode, the analysis finds a 1.9$\sigma$ exclusion of CP-conservation (defined as $\Jcp=0$) and a 1.2$\sigma$ exclusion of the inverted mass ordering.
\end{abstract}

\maketitle

\textit{Introduction}---Following the observation of neutrino oscillations \cite{PhysRevLett.81.1562}, experiments now aim to fully characterize the three-flavor mixing paradigm described by the Pontecorvo–Maki–Nakagawa–Sakata matrix using neutrinos from different sources~\cite{PhysRevD.110.012005, T2K:2023smv, DayaBay1, DayaBay2, PhysRevD.109.072014, PhysRevLett.120.071801}. Here, neutrino mixing is governed by three mixing angles ($\theta_{13}$, $\theta_{23}$, and $\theta_{12}$), two mass splittings ($\Delta m^2_{32}$ and $\Delta m^2_{21}$), and one Charge Parity (CP) violating phase (\deltacp). While some oscillation parameters have been precisely measured \cite{PhysRevD.98.030001}, others remain relatively unconstrained. In particular, the CP-violating phase, the ordering of the neutrino mass states (MO), and the octant of $\theta_{23}$ have not been determined experimentally. 
The magnitude of CP violation is proportional to the Jarlskog invariant~\cite{KRASTEV198884,PhysRevLett.55.1039,JARLSKOG2005323},
 $\Jcp = \sin\theta_{13}\cos^2\theta_{13}\sin\theta_{12}\cos\theta_{12}\sin\theta_{23}\cos\theta_{23}\sin\delta_{\scriptscriptstyle\mathrm{CP}}$.

\medskip

\textit{Experimental setup}---The Super-Kamiokande (SK) experiment \cite{Fukuda:2002uc} measures atmospheric neutrino oscillations using a large multi-purpose water Cherenkov detector located in the Kamioka mine in Gifu, Japan.  The detector has a 32~kiloton inner detector optically separated from a 2~meter thick outer detector, which mainly serves as a veto region.  Atmospheric neutrinos, produced by the interaction of cosmic rays with nuclei in the Earth's atmosphere, include a mixture of neutrino flavor states, as well as a wide range of propagation baselines (15\(\sim\)13000~km) and neutrino energies (MeV\(\sim\)TeV). 

The Tokai-to-Kamioka (T2K) long-baseline neutrino experiment \cite{Abe:2011ks} measures neutrino oscillations over a baseline of 295~km using a primarily muon (anti)neutrino beam produced by the neutrino facility at J-PARC, located in Ibaraki, Japan. SK is T2K's far detector (FD) and measures neutrinos after oscillations 2.5$^\circ$ off of the beam axis. The beam neutrino flux and neutrino interaction cross sections are constrained by a suite of near detectors (T2K ND) situated 280~m downstream of the neutrino production target.

\medskip

\textit{Motivation for a joint analysis}---T2K and SK have complementary strengths to study neutrino oscillations. T2K's off-axis neutrino beam provides a narrow energy spectrum peaked at 600~MeV and a known direction for beam-induced events at SK. This enables a precise measurement of the ``disappearance'' of $\numuany$ through oscillations, which manifest as a dip around 600~MeV in the spectra of $\numuany$ events observed at SK. 
T2K precisely measures $|\Delta m^2_{32}|$ and $\sin^2(2\theta_{23})$, which are connected to the peak energy and the amplitude of this disappearance, respectively. The MO and the value of $\sin(\deltacp)$ both affect the ``appearance'' probabilities for neutrinos P($\numu\rightarrow \nue$) and antineutrinos P($\numub \rightarrow \nueb$) asymmetrically. 
T2K's ability to change the beam composition from primarily neutrinos to primarily antineutrinos gives a powerful way to compare the oscillations of the two, enabling some constraint on \deltacp and the MO from the numbers of $\nue$ and $\nueb$ events observed at the far detector. 
The similarity of the effects of the MO and the value  of $\sindcp$ can however seriously degrade the experiment's ability to measure these parameters (Appendix~\ref{app:A}).

SK's atmospheric neutrino sample, on the other hand, provides a comparatively weak constraint on $|\Delta m^2_{32}|$ and $\sin^2(2\theta_{23})$, due to limited information about the incoming neutrino direction, and a broader range of neutrino energies. However, upward-going neutrinos experience large matter effects, which asymmetrically modify the oscillation probabilities of neutrinos and antineutrinos, dependent upon the MO. 
In particular, 2\(\sim\)10~GeV (anti-)neutrinos experience a resonant enhancement of their appearance probability if the ordering is normal (inverted). The size of this enhancement is proportional to $\sin^2(\theta_{23})$. This provides sensitivity to both the octant of $\theta_{23}$ and the MO through an excess of upward-going $\nue$ or $\nueb$ events at these energies. In a combined analysis, this provides a means of breaking the T2K MO--$\deltacp$ degeneracy, complementing the MO sensitivity achieved at T2K through its lower energy $\nueany$ appearance events. 
Additionally, systematic uncertainties at SK are better constrained in a joint fit than in the individual experiments: the beam and atmospheric samples at SK share numerous detector uncertainties and both can receive neutrino interaction uncertainty constraints from the T2K ND.

\medskip

\textit{Analysis strategy}---The analysis described here is based on previous analyses from the two experiments ~\cite{T2K:2023smv, Super-Kamiokande:2019gzr}, modified to produce a coherent joint analysis. Neutrino oscillation parameters are measured by comparing predictions for the rates and spectra of the atmospheric and beam neutrinos to observations performed at SK. The predictions are made using a model of the two experiments, covering neutrino fluxes, interactions, and detector response, with associated uncertainties. This model is built unifying aspects of the two experiments' analyses where appropriate, and using each individual experiment's approach otherwise.

Due to similarities in the neutrino energy spectra and event selections, T2K and low energy atmospheric events are described by a common neutrino interaction model. Largely independent interaction systematic uncertainties are used for higher energy atmospheric events as measurements from the T2K ND are not always applicable to these events, though the base interaction model is the same for all events.
The neutrino flux models for the experiments~\cite{T2K:2023smv,T2KFlux,Honda2011} are mostly independent, with the only common source of systematic uncertainty coming from hadron production in proton collisions. 
Hadron production is tuned using different, independent measurements in the two models: the
SK atmospheric flux model uses atmospheric muon measurements~\cite{TsukubaFlux, SummerFlux}, whereas T2K's model uses measurements by the NA61/SHINE experiment~\cite{Abgrall:2016jif}.
The neutrino events of the two experiments are observed in the same detector, and the correlated effects of detector systematic uncertainties on SK and T2K event samples were evaluated for the joint analysis.

\medskip

\textit{Event selection}---This analysis uses a total of eighteen SK atmospheric and five T2K event samples, constructed as described in Refs.~\cite{Super-Kamiokande:2019gzr} and \cite{T2K:2021xwb}. The event selections are based primarily on the number of reconstructed Cherenkov rings, the type of those rings, and the number of delayed Michel electron candidates. The ring types are either showering ($e$-like) or non-showering ($\mu$-like) and are the basis of the separation between \nueany and \numuany events.
The T2K selections target events with little activity in the SK outer detector, so-called fully contained (FC) events, with a single Cherenkov ring. 
This topology primarily selects charged current quasi-elastic (CCQE)-like events.
However, in neutrino running mode an additional sample probes \nue events containing a below-Cherenkov-threshold $\pi^{+}$ by requiring exactly one $e$-like ring and one Michel electron. 
Atmospheric neutrinos span a much wider range of energies, and
the atmospheric FC sample is divided into sub- and multi-GeV categories based on the deposited visible energy in the detector.
The SK analysis additionally includes events with significant energy deposition in the outer detector. 
The T2K beam samples and FC single-ring sub-GeV atmospheric samples in SK have a large kinematic overlap, but differ slightly in their respective event selections.
The selections remain unchanged relative to the publications above.
One additional selection criterion, however, is applied to all SK FC and T2K samples to remove neutron contamination from the Michel candidates for each event.
This cut changes the event rates by  $\mathcal{O}(1\%)$.

\medskip

\textit{Interaction model}---Neutrino interactions are simulated with the NEUT generator v5.4.0~\cite{Hayato:2021heg} using the same configuration as T2K's analysis~\cite{T2K:2023smv}. The common ``low-energy'' uncertainty model used for the T2K and atmospheric SK FC sub-GeV samples is based on T2K's model, with two additions to cover important uncertainties for the  atmospheric samples. Additional normalization uncertainties on the neutral current single $\pi^0$ model are introduced 
motivated by studies of MiniBooNE data~\cite{MiniBooNE:2009dxl,Stowell:2016jfr}.
These uncertainties separately scale the resonant and coherent components.
A supplementary uncertainty on the CCQE cross-section ratio $\sigma_{\nu_e}/\sigma_{\nu_\mu}$ is added based on the difference of this ratio between the spectral function model~\cite{Benhar:1994hw} used in this analysis and new calculations using the Hartree-Fock model with Continuum Random Phase Approximation~\cite{HF-CRPA, NueNumu}. 
This uncertainty changes the number of events in the e-like atmospheric sub-GeV sample targeting events without pions in the final state (CC0$\pi$) by 2.2\%, and is the interaction uncertainty with the largest impact on the atmospheric sample contribution to the \deltacp measurement.

The ability of the low energy model to describe the atmospheric sub-GeV samples is evaluated by comparing its predictions to the observed down-going data. 
Those events are mostly unaffected by oscillations and can therefore be used to test the model without biasing the oscillation measurement. 
Good agreement with data is found for the samples targeting CC0$\pi$ events when using the T2K ND constraint, while a small data excess is seen without it. 
For the samples targeting charged current single charged pion (CC1$\pi^+$) events, a significant data excess is seen: 225~events are observed in the $e$-like sample for 160.0 $\pm$ 12.6(stat) $\pm$ 14.3(syst) predicted using the T2K ND constraint and 84 events are observed in the $\mu$-like sample for 52.0 $\pm$ 7.2(stat) $\pm$ 5.3(syst) predicted. 
The T2K ND constraint reduces prediction in these samples by $\sim$20\% compared to the nominal prediction. 
The excess is localized at low lepton momentum in the $e$-like sample, and uniformly distributed in the $\mu$-like sample. 
An excess is also seen at low momentum in the corresponding $e$-like beam sample, but is not significant due to the low statistics of this sample.
As a result of the observed excesses, an interaction model uncertainty is added to change the shape of the pion three-momentum spectrum for charged-current resonant interactions by modifying the Adler angle~\cite{ADLER1968189} distribution, based on theoretically motivated~\cite{Rein:1980wg,Feynman:1971wr} and empirical modifications. A reconstruction uncertainty is also added, as detailed in the next section.

The model for the remaining SK samples (``high-energy model'') is based on that used in the SK analysis, with its pion secondary interaction and CCQE parts shared with the low-energy model. 
However, two high $Q^2$ (four-momentum transfer) CCQE normalization parameters are left uncorrelated with the low-energy model due to limited phase space overlap. 
The high energy model for pion final-state interactions is tuned to external data~\cite{elder}, as is done by T2K. 
Due to little overlap between the phase spaces of the near-detector and non-sub-GeV atmospheric samples, data from T2K's ND are used to constrain interaction uncertainties in the low-energy model and the correlated part of the high-energy model, excluding newly added uncertainties and other parts of the high-energy model.

\medskip

\textit{Detector model}---
Many of the detector uncertainties in the SK and T2K analyses are estimated from comparisons between atmospheric data and simulation. 
For these, correlated uncertainties are constructed by simultaneously evaluating the effects of detector parameter variations on the event rate in both the SK and T2K samples. 
Correlations between the reconstructed momentum scale uncertainties of the two experiments are found to have an impact on the \dmsqtwothree constraint obtained in the data fit. 
Other detector uncertainties from the reference analyses that are relevant for only one of the experiments are applied to the corresponding samples here.  
An additional systematic uncertainty is introduced for the sub-GeV samples targeting CC$1\pi^+$ events.
It allows single-ring single-Michel electron events with low lepton momentum to migrate between the \nue-like and \numu-like samples. The size of this uncertainty covers the excess in data observed for the down-going CC$1\pi^+$ \nue-like events at low momentum.

\medskip

\textit{Oscillation analysis}---Two Bayesian and two frequentist analyses were constructed (Appendix~\ref{app:fitter_details}). 
When two of those frameworks return different results for a given measurement, the more conservative option is reported.
For atmospheric oscillation probability calculations, path-dependent density averaging of  matter effects based on a four layer approximation of the PREM model~\cite{DZIEWONSKI1981297} is used and  fast atmospheric oscillations at low energy are smeared. The path-dependence yields more precise oscillation probabilities than the conventional approximation assuming layers of constant density. 
Reactor experiment measurements of $\theta_{13}$ using $\nueb$ disappearance, \ssqthtwoonethree$=0.0853 \pm 0.0027$~\cite{PhysRevD.98.030001, DayaBay1, DayaBay2, DoubleChooz, RENO}, are used as an external constraint. 

\medskip

\textit{Robustness studies}---Simulated datasets~\cite{T2K:2023smv}, generated using alternative models and fit using the nominal model are used to measure how $p$-values and oscillation parameter constraints would be affected if the assumed model is incomplete (Appendix \ref{app:FDS}).
Fourteen simulated datasets are considered, corresponding to alternative neutrino interaction models and data-driven effects at both T2K ND and SK. 
These studies are used to estimate, for example, how the observed atmospheric down-going CC$1\pi^+$ data excess could bias the results if it originated from an unknown systematic effect. 
Some of the simulated datasets produce a visible shift in the preferred values for $\Delta m_{32}^2$. 
The uncertainty on $\Delta m_{32}^2$ is therefore inflated by $3.6\times10^{-5 }\ \mathrm{eV^2}/c^4$ to account for these effects.

\medskip

\textit{Dataset}--- The atmospheric dataset is slightly increased compared to Ref.~\cite{Super-Kamiokande:2019gzr} to include the full Super-Kamiokande IV period (2008--2018), corresponding to a total live-time of 3244.4~days. The same T2K dataset as Ref.~\cite{T2K:2023smv} is used, corresponding to exposures of $19.7 \times 10^{20}$ and $16.3 \times 10^{20}$ protons on target in neutrino and antineutrino modes, respectively. 

\medskip

\begin{figure}[htbp]
    \centering
    \includegraphics[width=0.49\textwidth, trim={0mm 0mm 10mm 10mm}, clip]{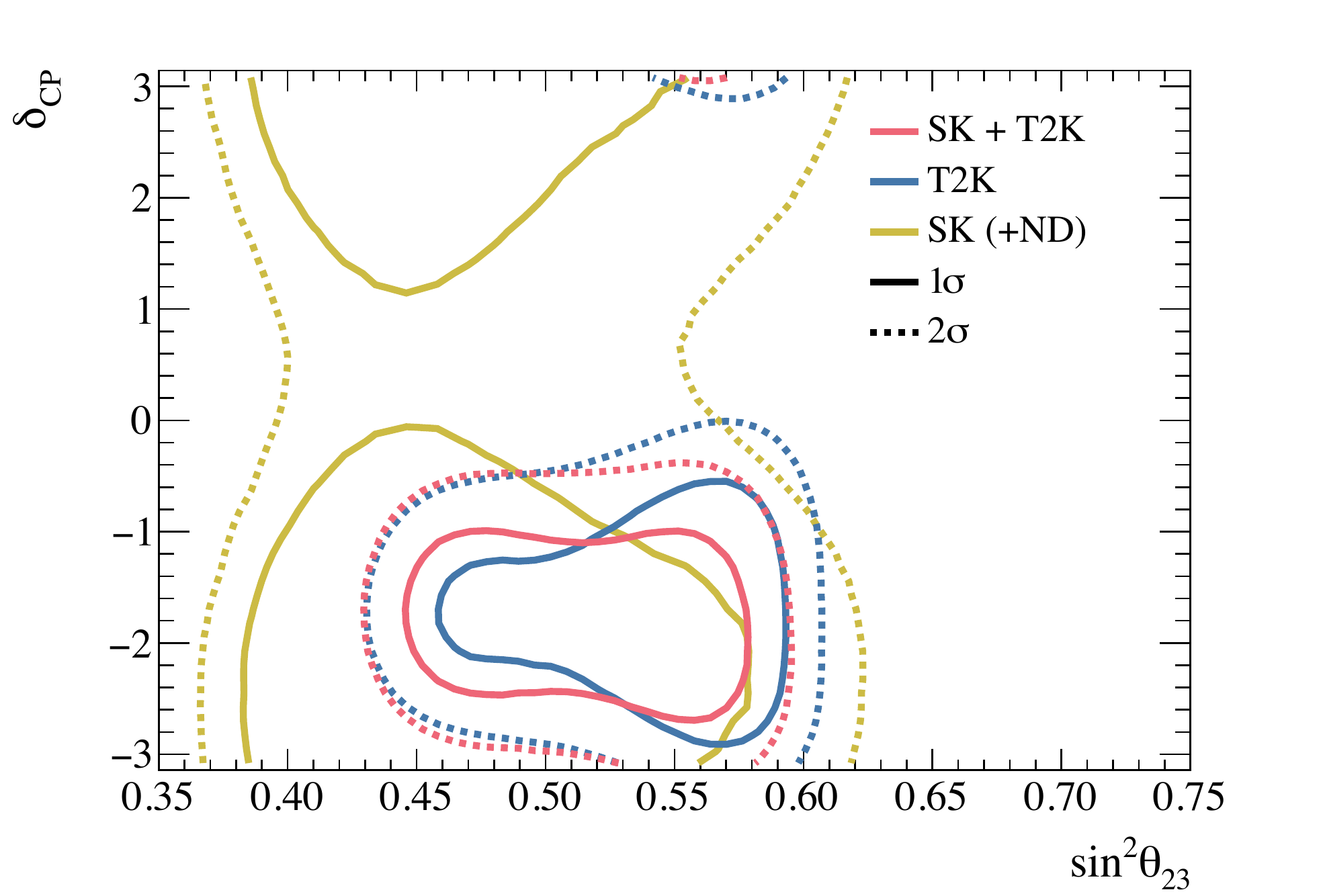} 
    \caption{The ($\ssqthtwothree$, $\deltacp$) credible regions obtained with the SK, T2K, and combined datasets.  The MO is marginalized over and a prior uniform in \deltacp is used.}
    \label{fig:Bayesian_sin2th23_dcp}
\end{figure}

\begin{figure}[hbtp]
    \centering
    \includegraphics[width=0.49\textwidth, trim={0mm 0mm 10mm 10mm}, clip]{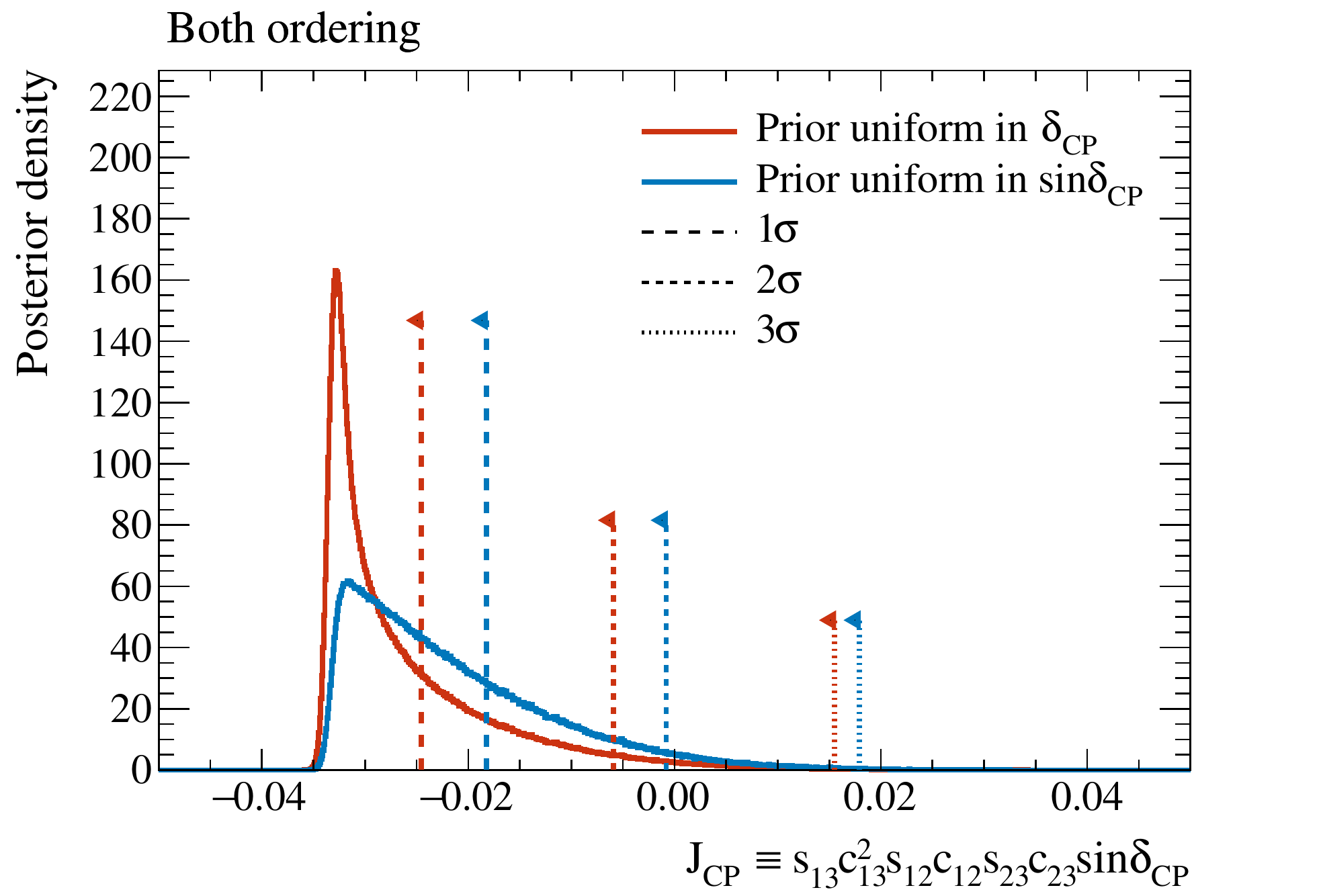}
    \caption{Posterior density for the Jarlskog invariant with credible intervals overlaid, marginalized over both MOs, and assuming a uniform prior in either \deltacp or \sindcp.
    The left edges of the intervals are close to each other in the region $-0.033 \leq \Jcp \leq -0.035$.}
    \label{fig:Bayesian_J}
\end{figure}

\textit{Bayesian results}---The Bayesian analyses assume uniform priors on \deltacp or \sindcp, \ssqthtwothree, $\Delta m^2_{32}$, and the MO. They find a preference for the normal ordering and a weak preference for the upper octant (\autoref{tab:BayesFactorsExp}).
SK and T2K data prefer different octants, which leads the joint analysis to have a weaker octant constraint than the individual experiments and a stronger preference for maximal mixing. 
Both experiments favor similar values of the CP-violating phase (\autoref{fig:Bayesian_sin2th23_dcp}).
The exclusion of CP-conserving values of \Jcp (\autoref{fig:Bayesian_J}) and \deltacp is reported as the largest fraction of the posterior density for which that value is not included in either of the two Bayesian analyses' highest posterior density credible intervals
(\autoref{tab:BayesianCP}).

\begin{table}[h]
\caption{\label{tab:BayesFactorsExp}
Octant and MO posterior probabilities using either the full dataset or samples from only one experiment and assuming equal prior probabilities. Values obtained by the second analysis are shown in parentheses.
}
\begin{ruledtabular}
\begin{tabular}{lccc}
 & SK only & T2K only & SK+T2K \\
\colrule
Upper octant    & 0.318 (0.337) & 0.785 (0.761) & 0.611 (0.639)\\
Normal ordering & 0.654 (0.633) & 0.832 (0.822) & 0.900 (0.887)\\
\end{tabular}
\end{ruledtabular}
\end{table}

\begin{table}[h]
\caption{\label{tab:BayesianCP}
Largest credible interval from the Bayesian analyses not containing different CP conserving values of $\Jcp$ and $\deltacp$. Values in parentheses indicate how these could change due to possible biases seen in robustness studies.
}
\begin{ruledtabular}
\begin{tabular}{l cc}
Value tested & \multicolumn{2}{c}{Prior uniform in} \\
 & $\deltacp$ & $\sin(\deltacp)$ \\
\colrule
$\Jcp=0$ & $2.3\sigma$ ($2.2\sigma$) & $2.0\sigma$ ($1.9\sigma$) \\
$\deltacp=0$ & $2.6\sigma$ ($2.5\sigma$) & $2.3\sigma$ ($2.2\sigma$)\\
$\deltacp=\pi$ & $2.1\sigma$ ($1.9\sigma$) & $1.6\sigma$ ($1.4\sigma$)\\
\end{tabular}
\end{ruledtabular}
\end{table}

\medskip

\begin{figure}[hbtp]
    \centering
    \includegraphics[width=0.49\textwidth, trim={0mm 0mm 0mm 10mm}, clip]{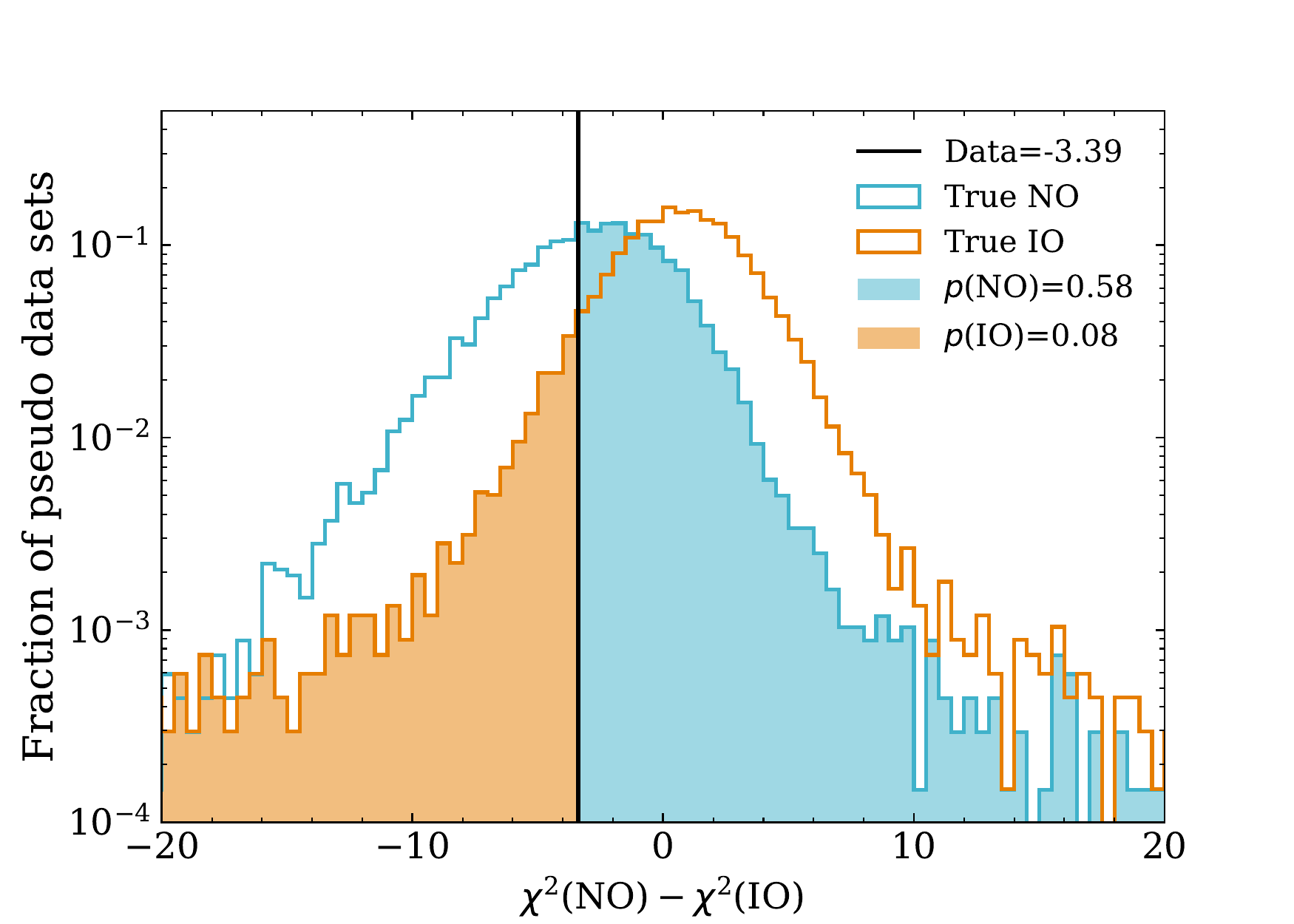}
    \caption{Distribution of the MO test statistic under true normal and inverted ordering hypotheses. The filled areas to the left (right) of the data result indicate the $p$-values for the inverted (normal) hypotheses.}
    \label{fig:MO}
\end{figure}

\textit{Frequentist results}--- The frequentist significance of the CP and MO results is evaluated using ensembles of pseudo-experiments. Estimating the significance of CP conservation (CPC) based on the presence or absence of both $0$ and $\pi$ in the \deltacp confidence intervals was found to have significant over-coverage.
Instead, the log-likelihood ratio between assuming CPC ($\sin\deltacp=0$, here equivalent to $\Jcp=0$) and without any assumption is used as a test statistic.
For the neutrino MO, the log-likelihood ratio between normal and inverted ordering is used (\autoref{fig:MO}).

The obtained $p$-values are summarized in \autoref{tab:Frequentist}. CPC is disfavored with a lower $p$-value ($p = 0.037$) than when using only the T2K data ($p = 0.047$).
Good agreement ($p = 0.75$) is found with an ensemble that allows for CP-violation by assuming posterior-distributed \deltacp values.
The inverted ordering is disfavored while good agreement with the normal ordering hypothesis is found, resulting in a $\mathrm{CL}_s$ parameter~\cite{CLs} for the inverted ordering of 0.18. 
The best-fit values and 68.3\% confidence intervals obtained using the Feldman--Cousins method~\cite{FC}  where necessary,
are $\deltacp = -1.76^{+0.73}_{-0.95}$, 
$\ssqthtwothree = 0.468^{+0.106}_{-0.025}$,
where MO was treated as a nuisance parameter,
and $\Delta m^2_{32}$ ($|\Delta m^2_{31}|$)${} = 2.520^{+0.048}_{-0.058}$ ($2.480^{+0.052}_{-0.048}$)${} \times 10^{-3} \,\mathrm{eV^2}$ for normal (inverted) ordering.

\begin{table}[h]
\caption{\label{tab:Frequentist}
Frequentist $p$-values for different CP and MO hypotheses. 
The most conservative of the two values obtained by the frequentist analyses is given. ``$p$-studies'' corresponds to the value up to which each $p$-value could increase due to biases seen in robustness studies.
}
\begin{ruledtabular}
\begin{tabular}{l c c}
Hypothesis & $p$-value & $p$-studies \\
\colrule
CP conservation & 0.037 & 0.050 \\
\colrule
Inverted ordering & 0.079 & 0.080 \\
Normal ordering & 0.58 & --- \\
\end{tabular}
\end{ruledtabular}
\end{table}

\medskip

\textit{Goodness of fit}---The Bayesian analyses find good posterior predictive $p$-values~\cite{Gelman_Post} using both the event spectra ($p=0.24$)
and total event counts ($p=0.19$).
The $p$-values for the individual T2K samples agree 
with the reference T2K analysis~\cite{T2K:2023smv} up to small differences coming predominantly from model changes.
The frequentist $p$-values~\cite{Maltoni:2003cu} additionally show consistency between the values of the systematic parameters favored by the T2K ND and atmospheric data ($p = 0.19$), as well as between the atmospheric and beam samples ($p = 0.24$). 

\medskip

\textit{Discussion}---The SK and T2K datasets favor similar values for the CP phase, close to maximal CP violation, and both show a preference for the normal MO. As a result, the combined analysis finds increased preferences for CP non-conservation and the normal ordering.
When looking directly at the exclusion of CPC through the presence of $\Jcp=0$ in credible intervals or frequentist $p$-values, an exclusion at the 2$\sigma$ level is found. 
However, the significance can fall below 2$\sigma$ due to potential weaknesses of the uncertainty model tested using simulated datasets. 
The alternative model assuming that the down-going CC1$\pi^+$-like data excess is completely due to an unknown systematic effect and the alternative nuclear model for CCQE interactions~\cite{PhysRevC.83.045501} 
have the largest impact.
The $p$-value obtained for the inverted ordering is significantly larger than the one obtained in the reference atmospheric analysis ($p=0.033$)~\cite{Super-Kamiokande:2019gzr}, despite the addition of the T2K samples which also favor the normal ordering. 
This was found to come mostly from the different values of $\ssqthtwothree$ favored by the two analyses (Appendix \ref{app:E}).

The constraint from the T2K ND has a small effect on the measurement of \deltacp and the MO using atmospheric samples. The most important systematic uncertainties for this \deltacp measurement are related to the atmospheric neutrino flux below 1 GeV, the $\sigma_{\nu_e}/\sigma_{\nu_\mu}$ cross-section ratio and the particle identification for single ring events, which cannot be directly constrained using the T2K ND. For the MO, most of the atmospheric neutrino sensitivity comes from the high energy samples where this constraint is not used for most interaction modes.
Accordingly, the MO sensitivity benefits mainly from the stronger 
$\ssqthtwothree$ constraint provided by T2K data.

\medskip

\textit{Conclusion}---The SK and T2K collaborations have produced a first joint analysis of their data. Common neutrino interaction and detector models have been developed for events from the two experiments with overlapping energies  and are found to properly describe both datasets. 
The results show an exclusion of the CP-conserving value of the Jarlskog invariant with a significance between 1.9$\sigma$ and 2.0$\sigma$, a limited preference for the normal ordering with a 1.2$\sigma$ exclusion of the inverted ordering~\cite{Conversion}, and no strong preference for the $\theta_{23}$ octant. This first joint analysis is an important step towards the combined beam and atmospheric data analyses planned by next-generation neutrino oscillation experiments.

\medskip

The data related to this work can be found in Ref.~\cite{ReleaseZenodo}.

\section*{Acknowledgements}
The Super-Kamiokande collaboration gratefully acknowledges cooperation of the Kamioka Mining and Smelting Company.
The Super-Kamiokande experiment was built and has been operated with funding from the
Japanese Ministry of Education, Science, Sports and Culture, 
and the U.S. Department of Energy.

The T2K collaboration would like to thank the J-PARC staff for superb accelerator performance. We thank the CERN NA61/SHINE Collaboration for providing valuable particle production data. We acknowledge the support of MEXT,   JSPS KAKENHI  and bilateral programs, Japan; NSERC, the NRC, and CFI, Canada; the CEA and CNRS/IN2P3, France; the DFG, Germany; the NKFIH, Hungary; the INFN, Italy; the Ministry of Science and Higher Education (2023/WK/04) and the National Science Centre (UMO-2018/30/E/ST2/00441, UMO-2022/46/E/ST2/00336 and UMO-2021/43/D/ST2/01504), Poland; the RSF (RSF 24-12-00271) and the Ministry of Science and Higher Education, Russia; MICINN and ERDF funds and CERCA program, Spain; the SNSF and SERI, Switzerland; the STFC and UKRI, UK; the DOE, USA; and NAFOSTED (103.99-2023.144, IZVSZ2.203433), Vietnam. We also thank CERN for the UA1/NOMAD magnet, DESY for the HERA-B magnet mover system, the BC DRI Group, Prairie DRI Group, ACENET, SciNet, and CalculQuebec consortia in the Digital Research Alliance of Canada, GridPP and the Emerald High Performance Computing facility in the United Kingdom, and the CNRS/IN2P3 Computing Center in France. In addition, the participation of individual researchers and institutions has been further supported by funds from the ERC (FP7), “la Caixa” Foundation, the European Union’s Horizon 2020 Research and Innovation Programme under the Marie Sklodowska-Curie
grant; the JSPS, Japan; the Royal Society, UK; French ANR and Sorbonne Université Emergences programmes; the VAST-JSPS (No. QTJP01.02/20-22);  and the DOE Early Career programme, USA. For the purposes of open access, the authors have applied a Creative Commons Attribution license to any Author Accepted Manuscript version arising.

\appendix
\section{\deltacp--MO degeneracy and the ability to reject CP conservation}
\label{app:A}

\begin{figure}[hbtp]
    \centering
    \includegraphics[width=0.49\textwidth]{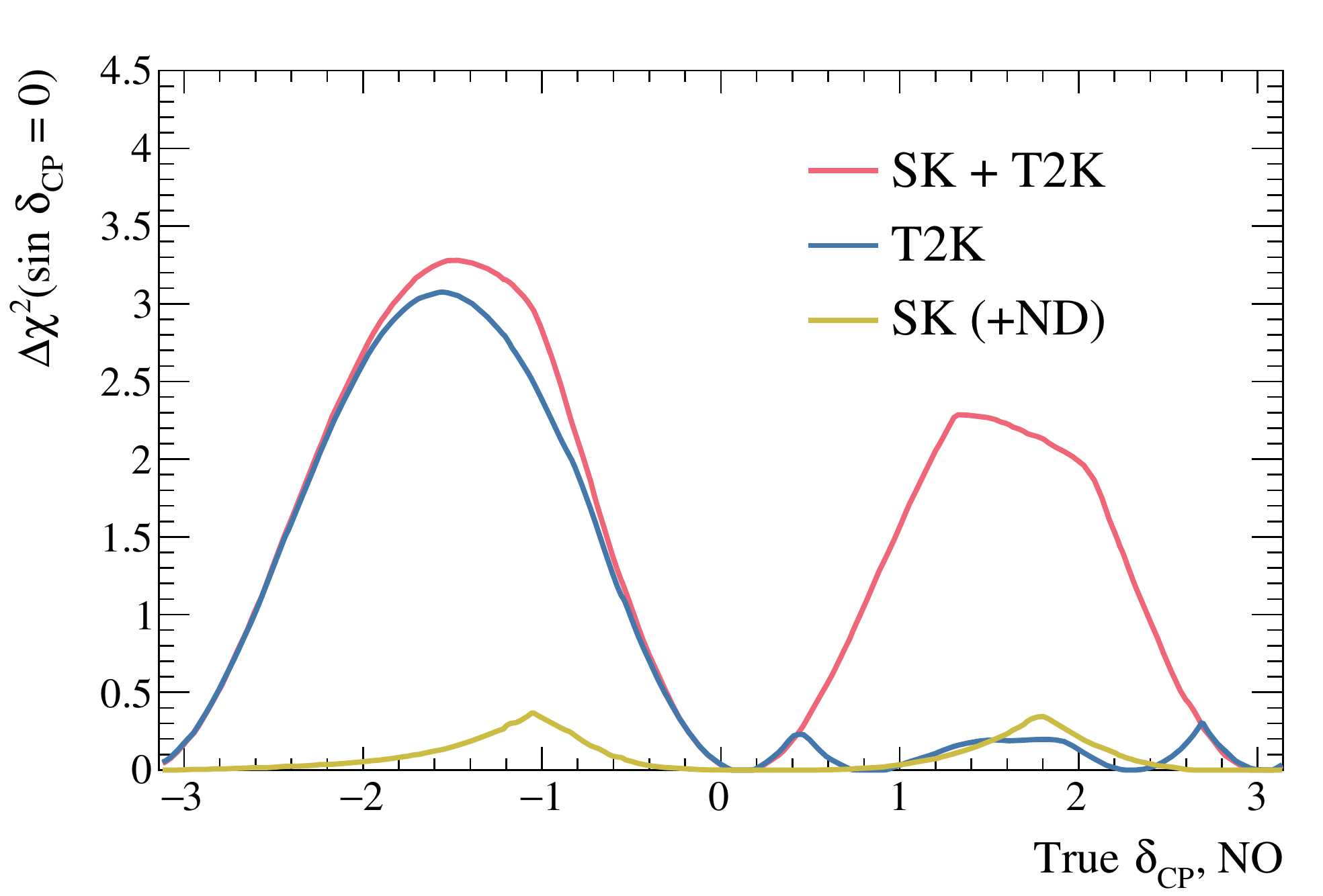}
    \caption{
    Sensitivity to reject the CP-conserving hypothesis for different true values of \deltacp assuming the normal MO. Other oscillation parameters are set to $\ssqthtwothree = 0.528$, $\ssqthonethree = 0.0218$, and $\dmsqtwothree = 2.509 \times 10^{-3} \, \mathrm{eV^2}$.}
    \label{fig:sensitivity:CPC}
\end{figure}

The impact of the \deltacp--MO degeneracy on rejecting CP conservation
is illustrated in Fig.~\ref{fig:sensitivity:CPC} assuming normal ordering.
T2K can reject the CP-conserving hypothesis if $\sindcp < 0$, which aligns with current measurements from T2K.
However, if $\deltacp\sim\pi/2$ and the mass ordering is normal---as weakly favored by NOvA data~\cite{NOvA:2021nfi,PhysRevD.110.012005}---T2K is largely insensitive to \deltacp due to the \deltacp--MO degeneracy, demonstrated in Fig. 18 of Ref.~\cite{T2K:2023smv}.
The degeneracy is resolved by SK's MO-constraint being decoupled from its \deltacp measurement in the joint analysis.
This results in a dramatic improvement in the joint analysis sensitivity compared to each experiments' individual sensitivities.
If the mass ordering is inverted, the loss of ability to reject CP conservation by T2K alone happens instead for $\deltacp\sim-\pi/2$, and the joint analysis significantly improves sensitivity in this region for the same reason.
These features were also confirmed by studying the statistical power to reject CPC using ensembles of pseudo datasets.

\section{Resonant and deep inelastic interaction uncertainty model}
In this analysis the uncertainty model for resonant and deep inelastic interactions for the low energy samples is based on the T2K model, while the model for the high energy samples is based on the SK model.
These two uncertainty models have many similarities. For resonant interactions, both use uncertainties on the axial mass, the axial form factor at $Q^2=0$ and the normalization of the non-resonant isospin-1/2 component and have similar implementations. The reference T2K and SK analyses use different prior uncertainties for these parameters. For the joint analysis presented here, prior uncertainties from the reference T2K analysis~\cite{T2K:2023smv} are used for all samples, which gives increased uncertainties compared to the SK reference analysis. 
Further, the T2K analysis includes an uncertainty on the normalization of the non-resonant background at low pion momenta for antineutrinos, while the SK analysis has additional uncertainties on the $\nu/\nubar$ and $1\pi^{0}/1\pi^{\pm}$ cross-section ratios for resonant interactions.
The latter are based on comparisons between the predictions of the nominal Rein--Sehgal model~\cite{Rein:1980wg} and those of the model by Hernandez et al.~\cite{PhysRevD.76.033005}.

For deep inelastic (DIS) interactions, the T2K and SK uncertainty models also have similarities as the T2K model is based on an older version of the SK one. 
Both models include uncertainties on the Bodek--Yang model~\cite{BY}. 
For the low invariant mass ($W<2$ GeV/c$^{2}$) region, this uncertainty is separated into uncertainties on the axial and vector parts and an extra normalization on the structure function from Ref.~\cite{bodek2021inelasticaxialvectorstructure} in the SK analysis.
In the T2K analysis a simple comparison between models with and without the Bodek--Yang corrections in this region is used. 
Both models include uncertainties on the number of hadrons produced in DIS interactions for the low $W$ region and use similar implementations. 
The SK analysis includes an additional normalization uncertainty based on comparisons between the predictions of the Bodek--Yang model and an alternative model~\cite{CKMT}.

\section{Technical details on the four oscillation analyses}
\label{app:fitter_details}
The analysis leverages results from four different frameworks; two using Bayesian (BA1, BA2) and two using frequentist (FA1, FA2) statistics.
BA1 and BA2 are based on T2K analyses and FA1 is a modified version of BA1 optimized to make frequentist analyses computationally tractable. 
The second frequentist analysis is based on the SK analysis and acted also as a validation of the implementation of the SK experiment in the T2K-based frameworks.
Similarly, the implementation of the T2K experiment in FA2 was validated using comparisons to the T2K-based analyses.
The four frameworks made different, valid choices for treating systematic uncertainties, calculating oscillation probabilities in matter, binning observables, and implementing statistical methodology. 
Analysis conclusions were largely invariant to these choices.

Both BA1 and BA2 use Markov Chain Monte Carlo (MCMC) methods to evaluate the marginal likelihoods for the parameters of interest. 
On the other hand, FA1 and FA2 compute the profile likelihood on a fixed grid of the oscillation parameters of interest. 
All analyses utilize a binned negative log-likelihood test-statistic, assuming Poisson-distributed statistics and Gaussian penalties when systematic parameters vary away from their nominal values.

In T2K's analyses external data and the ND are used to constrain the majority of systematic uncertainties improving the FD samples' sensitivity to neutrino oscillation parameters. 
BA2 performs a simultaneous fit of the T2K ND, T2K FD, and SK atmospheric data, while the other analyses use a covariance matrix to encode the ND constraint. 
The covariance matrix approach assumes a Gaussian probability density for the systematic uncertainties, whereas the direct implementation of the ND in BA2 avoids this~\cite{T2K:2023smv}. 
The large number of samples and systematic parameters in the joint analysis makes fits of ensembles of pseudo datasets computationally challenging. 
FA1 therefore assumes linearized response functions for the systematic parameters to be able to use a faster minimization method.
BA1, FA1 and FA2 assumes the energy scale for atmospheric and beam events as completely correlated, while BA2 considers them uncorrelated.

The T2K samples are binned in combinations of \textit{reconstructed} lepton momentum, $p_\textrm{lep}$, angle with respect to the beam, $\theta_\textrm{lep}$, and neutrino energy, $E_\nu^\textrm{rec}$~\cite{T2K:2023smv}.
For the $e$-like samples, BA1, FA1 and FA2 use $(p_\textrm{lep},\theta_\textrm{lep})$ while BA2 uses $(E_\nu^\textrm{rec},\theta_\textrm{lep})$. 
For the $\mu$-like samples, BA1 and FA2 use $(E_\nu^\textrm{rec},\theta_\textrm{lep})$, and BA2 and FA1 only use $E_\nu^\textrm{rec}$ information to decrease computational overhead, with a slight loss in sensitivity to $\Delta m^2_{32}$. 
For the SK samples, all analyses use the sample definitions and binning in visible energy and zenith angle from the reference analysis~\cite{Super-Kamiokande:2019gzr}.
Fast oscillations at low energy in the atmospheric samples are smeared by semi-analytic averaging in BA1 and FA1, binned down-sampling in BA2, and binned neighbor smearing in FA2. 
Both FA1 and FA2 place \deltacp, \ssqthtwothree, \dmsqtwothree and the mass ordering on a fixed grid while fitting nuisance parameters, while the Bayesian frameworks calculate the oscillation probability on-the-fly.

\medskip

\section{Simulated data studies}
\label{app:FDS}
The ``simulated data study" method is used in T2K to test the robustness of the analysis' results with respect to systematic effects not explicitly implemented in the uncertainty model. 
The T2K approach to oscillation analysis uses an uncertainty model which is constrained using data observed at the ND. However, if an important systematic effect is not part of the systematic parameters used in the T2K ND analysis, the fit could nevertheless find a combination of parameters that makes predictions agree with data.
In such a case, the extrapolation of the tuned model to the far detector could yield incorrect predictions due to the difference of detector acceptance and neutrino fluxes (from oscillations), which could then bias the measurements of neutrino oscillations.

There are currently no neutrino interaction models that can describe all available neutrino scattering data simultaneously. A number of models exist which perform equally well when compared to these data but differ in their mapping of the neutrino energy, which is relevant for oscillations, to the observables in the detectors. These models do not necessarily cover all possible unknown effects, and the analysis needs to be robust to a whole range of possible model variations. In the main analysis, one baseline model is chosen and extended to allow systematic variations. When another model is not explicitly used in the construction of the uncertainty model, it can be used as an example of a plausible deviation from the baseline model to test for potential weaknesses of the systematic uncertainty model. In addition, ad-hoc alternative models can be created to test the effect of variations in other parts of the model.
In practice datasets are created at the near and far detectors using an alternative model but then fitted assuming the nominal model. 
Oscillation parameter measurements are then compared to the ones obtained from fitting a dataset based on the nominal model to estimate the possible bias from the effect tested.
More detailed explanations of the procedure can be found in Refs.~\cite{T2K:2023smv} and \cite{PhysRevD.96.092006}.

Fourteen such simulated data studies were performed for the analysis described in this letter. Most of the alternative models considered are related to neutrino interactions, being either alternative models for a given type of interaction, alternative nuclear models, or data-driven effects. The first group of alternative models tested corresponds to those that had the most impact in the reference T2K analysis~\cite{T2K:2023smv}. They correspond to alternative models for 2p2h interactions~\cite{PhysRevC.80.065501} and the axial form factor for CCQE interactions (using a three component extension of Ref.~\cite{PhysRevC.78.035201}), a change in the value of the binding energy for CCQE interactions and data-driven effects from the differences between predictions and data at the T2K ND. Alternative nuclear models~\cite{PhysRevC.83.045501, CRPA} used in simulated data studies in more recent versions of the T2K analysis are also considered, as well as additional effects deemed relevant for the combined analysis, specifically alternative models for hadron production in DIS interactions~\cite{doi:10.7566/JPSCP.12.010041} and energy dependence of the CCQE cross-section ratio $\sigma_{\nu_e}/\sigma_{\nu_\mu}$. Finally, the procedure is used to check whether the data excess observed in the atmospheric down-going CC$1\pi^+$ samples could significantly affect the results if it was due to an unknown deficiency of the model.

\section{Generation of pseudo datasets for frequentist results}
\label{app:E}
Ensembles of pseudo datasets are constructed
to evaluate the frequentist significance of the CP and MO results, taking into account statistical fluctuations and randomizing the values of nuisance oscillation and systematic parameters according to their posterior~\cite{Cousins:1991qz} and prior probability distributions, respectively. 
Even after profiling or marginalizing over these parameters, the distributions of the obtained test-statistics retain some dependence on the parameter values assumed in generating the datasets.
The distribution of the MO test statistic, in particular, depends on the assumed values of \ssqthtwothree (from SK and T2K samples) and \deltacp (from T2K samples). 
For SK samples the dependence arises because the magnitude of the MO-sensitive matter resonance is proportional to \ssqthtwothree. 
For the T2K samples the almost identical impact of changes in the MO and shifts in $\sindcp$ on  \nue/\nueb-appearance probabilities gives rise to a strong dependence on the assumed value of \deltacp.

In the joint fit, the dependence on \deltacp is reduced by the independent MO-constraint from the atmospheric samples. 
A sub-leading dependence on \ssqthtwothree also exists in T2K, since its symmetric impact on the \nue/\nueb-appearance probability is not easy to distinguish from the anti-symmetric impact of MO due to limited \nueb statistics.
For the joint fit the $p$-value for the inverted ordering varies between 0.05 and 0.08 when assuming different true values for \ssqthtwothree and \deltacp over the range of their 90\% confidence intervals.
This dependence causes a noticeable change in the strength of the MO-constraint from the atmospheric samples compared to the baseline SK analysis~\cite{Super-Kamiokande:2019gzr}: since the T2K samples pull \ssqthtwothree toward the upper octant, the SK constraint on the MO becomes weaker and comparable to that from T2K.

\bibliography{main}

\begin{thebibliography}{51}
\bibitem [{Con()}]{Conversion}%
  \BibitemOpen
  \href@noop {} {}\bibinfo {note} {For the z-score calculation, the mass
  ordering test is treated as a one-sided test (1$\sigma$ corresponds to a
  84.1\% probability), contrary to the one for CP symmetry which is treated as
  a two-sided test (1$\sigma$ corresponds to a 68.3\% probability)}\BibitemShut
  {NoStop}%
\end{thebibliography}

\end{document}